\documentclass[%
reprint,
superscriptaddress,
showkeys,
 amsmath,amssymb,
 aps,prl
]{revtex4-2}

\usepackage{graphicx}
\usepackage{dcolumn}
\usepackage{bm}
\usepackage{xcolor}

\usepackage[normalem]{ulem}
\usepackage{cancel}

\begin{document}

\preprint{}

\title{A physics-aware machine to predict extreme events in turbulence}

\author{N.A.K. Doan}
\affiliation{%
 Department of Mechanical Engineering, Technical University of Munich, Garching 85747, Germany}%
\affiliation{%
 Institute for Advanced Study, Technical University of Munich, Garching 85748, Germany}%
\author{W. Polifke}%
\affiliation{%
 Department of Mechanical Engineering, Technical University of Munich, Garching 85747, Germany}%

\author{L. Magri}\email{lm547@cam.ac.uk}
\affiliation{
 Department of Engineering, University of Cambridge, Cambridge CB2 1PZ, United Kingdom
}%
\affiliation{%
 Institute for Advanced Study, Technical University of Munich, Garching 85748, Germany (visiting)}%

\date{\today}
\begin{abstract}
We propose a physics-aware machine learning method to time-accurately predict extreme events in a turbulent flow.  
The method combines two radically different approaches: empirical modelling based on reservoir computing, which learns the chaotic dynamics from data only, and physical modelling based on conservation laws. We show that the combination of the two approaches is able to predict the occurrence and amplitude of extreme events in the self-sustaining process in turbulence---the abrupt transitions from turbulent to quasi-laminar states---which cannot be achieved by using either approach separately. 
This opens up new possibilities for enhancing synergistically data-driven methods with physical knowledge for the accurate prediction of extreme events in chaotic dynamical systems. 
\end{abstract}
\maketitle
{\it Introduction.--} 
%
 %
 Extreme events occur in many natural and engineering systems~\cite{Farazmand2019}, such as oceanic rogue waves~\cite{Dysthe2008}, extreme climate and weather events, e.g., flooding and storm damage \cite{Easterling2000,Majda2012}, intermittency in turbulence~\cite{Platt1991}, and thermoacoustic instabilities in aeroengines and rocket motors~\cite{Lieuwen2006a}, to name only a few. In this Letter, we focus on abrupt self-sustaining process events in a turbulent flow~\cite{Moehlis2004}. Turbulent flows are chaotic dynamical systems that are extremely sensitive to small perturbations to the system. This is commonly referred to as the {\it butterfly effect}~\cite{Lorenz1963} in chaos theory. Because of the butterfly effect, the time accurate prediction of chaotic flows can only be achieved for a typically short time, which is called the predictability time.  
This is a roadblock for the {\it time-accurate} prediction of extreme events because, after the predictability time, a minuscule difference between the initial conditions, such as floating-point errors, is exponentially amplified. Because of this, the time-accurate prediction of extreme events is still an open problem~\cite{Farazmand2019}. 
The state-of-the-art in the prediction of extreme events chiefly relies on statistical approaches, e.g., Extreme Value Theory~\cite{Nicodemi2012}  and Large Deviation Theory~\cite{Varadhan2008}. These methods characterize the probability of the occurrence of an event and the heavy tail of the probability density function of the observable associated with the event. Notably, Sapsis~\cite{Sapsis2018} combined Large Deviation Theory with data-driven methods to characterize efficiently the heavy tail of the distribution. These statistical methods provide an excellent framework to identify precursors and calculate the probability of extreme events, but they do not provide a robust way to time-accurately predict their occurrence and amplitude.  
Recently, machine learning and data-driven methods have shown great potential in learning the unpredictable dynamics of chaotic systems. 
In particular, Echo State Networks~\cite{Jaeger2007,Lukosevicius2009} (ESNs), which are a class of recurrent neural networks based on reservoir computing, have proved successful in learning the chaotic dynamics beyond the predictability time~\cite{Pathak2018,Pathak2018a,Doan2019}.  
ESNs predict the dynamics of chaotic systems by learning temporal patterns in data only, but the learned solutions may violate physical principles. Turbulent flows, however, must obey physical principles such as momentum and mass conservation. The over-reaching objective of this Letter is to propose a machine learning method that produces physical solutions to predict extreme events in a turbulent flow. We show that constraining the physical principles in the training of the machine is key to the time accurate prediction of an extreme event. 

\begin{figure}[!ht]
    \centering
    \includegraphics[width=7cm]{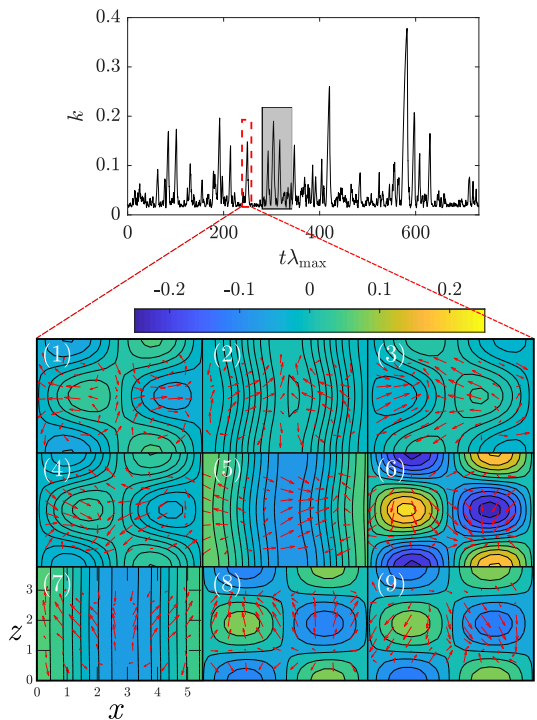}
    \caption{Kinetic energy, $k$, and velocity field in the mid-$y$ plane. The arrows indicate the in-plane velocity ($x$-$z$ directions), the coloured contour indicates the ouf-of-plane velocity, and the grey box indicates the data used for the training of the PI-ESN.%
    }
    \label{fig:MFE_evol}
\end{figure}

{\it Turbulent flow model.--} 
 To describe the self-sustaining process in turbulence, we regard the turbulent flow as an autonomous dynamical system
 $\dot{\bm{y}} = \mathcal{N} (\bm{y})$ with $\bm{y}(0)  =\bm{y}_0$, where 
$\dot{(\;)}$ is the temporal derivative; and $\mathcal{N}$ is a deterministic nonlinear  differential operator, which encapsulates the numerical discretization of the spatial derivatives and boundary conditions (if any). 
The turbulent flow under investigation is incompressible. The domain is a cuboid of size $L_x \times L_y \times L_z$ between two infinite parallel walls at $y=0$ and $y=L_y$, which are periodic in the $x$ and $z$ directions. A sinusoidal volume force is applied in the $y$-direction. The flow is governed by momentum and mass conservation laws, i.e., the Navier-Stokes equations, which were reduced in form by Moehlis, Faisst and Eckhart (MFE) \cite{Moehlis2004} (see Supplementary Material). The MFE model, which provides the operator $\mathcal{N}$ in the dynamical system formulation, captures the essential features of the transition from turbulence to quasi-laminar states such as the exponential distribution of turbulent lifetimes. The velocity field is decomposed as $\bm{v}({\bm x},t) = \sum_{i=1}^9 a_i (t) \bm{v_i}(\bm{x})$, where $\bm{v_i}(\bm{x})$ are spatial Fourier modes (or combinations of them)~\cite{Moehlis2004}. Hence, the Navier-Stokes equations are projected onto $\bm{v_i}(\bm{x})$ to yield nine ordinary differential equations for the modes' amplitudes, $a_i$, which are nonlinearly coupled. Consequently, the state vector is ${\bm y}=\{a_i\}_1^9$. All the variables are non-dimensional~\cite{Moehlis2004}.  
%
%
Physically, 
$\bm{v}_1$ is the laminar profile mode; 
$\bm{v}_2$ is the streak mode; 
$\bm{v}_3$ is the downstream vortex mode; 
$\bm{v}_4$ and $\bm{v}_5$ are the spanwise flow modes; 
$\bm{v}_6$ and $\bm{v}_7$ are the normal vortex modes; 
$\bm{v}_8$ is the three-dimensional mode; and 
$\bm{v}_9$ is the modification of the mean profile caused by turbulence.
The flow has a fixed point $a_1=1$, $a_2=...=a_9=0$, which is a laminar state~\footnote{The Supplementary material reports the initial condition, the expressions for $\bm{v_i}$ and the equations for $a_i$~\cite{Moehlis2004}.
Detailed analysis of the MFE model is available in \cite{Kim2008,Joglekar2015}. }. 
The domain size is $L_x=1.75\pi$, $L_y=2$ and $L_z=1.2\pi$. The Reynolds number is $600$. The initial condition is such that the turbulent flow has chaotic bursts between fully turbulent and quasi-laminar states. These are the extreme turbulent events we wish to predict.
The governing equations are integrated in time with a 4$^{th}$-order Runge-Kutta scheme~\cite{Press1992} with a time step $\Delta t = 0.25$. This provides the evolution of the nine modes $a_i$ from $t=0$ to  $t=30,000$  (Fig.~\ref{fig:MFE_evol}, top panel).
The evolution of the kinetic energy, $k= 0.5 \sum_{i=1}^9 a_i^2$, is shown in Fig.~\ref{fig:MFE_evol}. The time is normalized by the largest Lyapunov exponent, $\lambda_{\max}=41$, which was calculated as the average logarithmic error growth rate between two nearby trajectories~\cite{Boffetta2002}. (The Lyapunov time scale is $\lambda_{\max}^{-1}$.)
The kinetic energy, $k$, has sudden large peaks, which suddenly burst from smaller chaotic oscillations. 
%
Each burst is a quasi-relaminarization event, which occurs in three phases (Fig.~\ref{fig:MFE_evol}):  
(i) the originally laminar velocity profile becomes unstable and breaks down into vortices due to the shear imposed by the volume force (panels 5-7);
(ii) the vortices align to become streaks (panels 8-9 and 1-2); and
(iii) the streaks break down leading to flow relaminarization (panels 3-5). 

\begin{figure}[!ht]
    \centering
    \includegraphics[width=7cm]{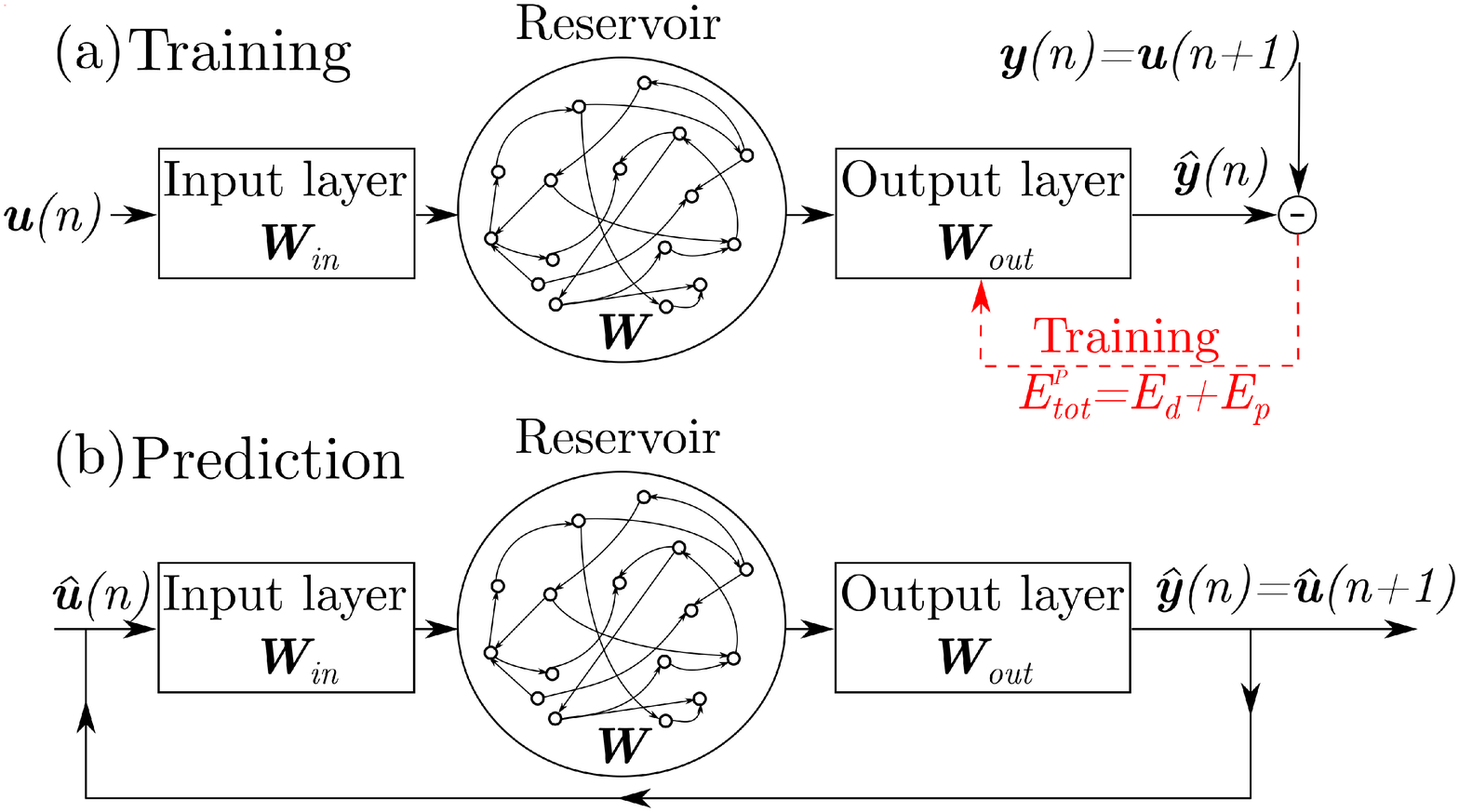}
    \caption{PI-ESN during (a) training and (b) prediction.}
    \label{fig:ESN_schema}
\end{figure}

{\it Physics-aware reservoir computing.--}
To learn the turbulent dynamics, we constrain the physical knowledge of the turbulent flow into a reservoir computing data-driven method based on the Echo State Network~\cite{Jaeger2007,Lukosevicius2009} (ESN): The Physics-Informed Echo State Network~\cite{Doan2019} (PI-ESN). A schematic is shown in Fig.~\ref{fig:ESN_schema}. 
We have training data with an input time series $\bm{u}(n)\in \mathbb{R}^{N_u}$ and a target time series $\bm{y}(n)\in \mathbb{R}^{N_y}$, where $n=0,1,2,\ldots, N_t$ are the discrete time instants that span from $0$ to $T=N_t\Delta t$. During prediction, the target at time $n$ becomes the input at time $n+1$, i.e., $\bm{u}(n+1)=\bm{y}(n)$. 
The training of the PI-ESN is achieved by (i) minimizing the error between the prediction, $\widehat{\bm{y}}(n)$, and the target data ${\bm{y}}(n)$ when the PI-ESN is excited with the input, $\bm{u}(n)$ (Fig. \ref{fig:ESN_schema}a), and (ii) enforcing that the prediction does not violate the physical constraints. To enforce (ii), we observe that 
a solution of the turbulent flow, $\bm{y}=\{a_i\}_1^9$, is such that the {\it physical error} (also known as the residual) is zero, i.e., $\mathcal{F}(\bm{y})\equiv \dot{\bm{y}}-\mathcal{N}(\bm{y})=0$.  %
To estimate the physical error beyond the training data, the PI-ESN is looped back to its input (Fig.~\ref{fig:ESN_schema}b) to obtain predictions $\lbrace \widehat{\bm{y}} (n_p)  \rbrace_{p=1}^{N_p}$ in the time window $(T+\Delta t) \leq t \leq (T+N_p\Delta t)$. The number of collocation points, $N_p$, is user-defined.  The physical error $\mathcal{F}({\widehat{\bm{y}}(n_p)})$ is evaluated to train the PI-ESN such that the sum of 
(i) the physical error between the prediction and the available data from $t=0$ to $t=T$, $E_d$, and 
(ii) the physical error for $t>T$, $E_p$, is minimized. Mathematically, we wish to find $\widehat{{\bm y}}(n)$ for $n=0,1,\ldots,N_t+N_p$ that minimizes 
\begin{multline}
E_{tot}^P =  \underbrace{\frac{1}{N_t} \sum_{n=1}^{N_t} \lvert\lvert\widehat{{\bm y}} (n) - {\bm y} (n)\lvert\lvert^2 }_{E_{d}} + \underbrace{  \frac{1}{N_p} \sum_{p=1}^{N_p} \lvert\lvert \mathcal{F}(\widehat{{\bm y}}(n_p))\lvert\lvert^2 }_{E_p}, 
\label{eq:EPhys}
\end{multline}
where $\lvert\lvert\cdot\lvert\lvert$ is the Euclidean norm. Note that the PI-ESN is straightforward to implement because it is requires only cheap residual calculations at the collocation points, i.e., it does not require solving for the exact solution. 
%
%
The architecture of the PI-ESN follows that of the ESN, which consists of an input matrix $\bm{W}_{in}\in\mathbb{R}^{N_x \times N_u}$, which is a sparse matrix; a \textit{reservoir} that contains $N_x$ neurons that are connected by the  recurrent weight matrix $\bm{W}\in\mathbb{R}^{N_x \times N_x}$, which is another sparse matrix; and the output matrix $\bm{W}_{out}\in\mathbb{R}^{N_y\times N_x}$. The input time series, $\bm{u}(n)$, is connected to the reservoir through $\bm{W}_{in}$ to excite the states of the neurons, $\bm{x}$, as 
$
    \bm{x}(n+1) = \tanh \left( \bm{W} \bm{x}(n) + \bm{W}_{in} \bm{u}(n+1) \right)
$, 
where $\tanh(\cdot)$ is the activation function. The output of the PI-ESN, $\widehat{\bm{y}}(n)$, is computed by linear combination of the reservoir states as $\widehat{\bm{y}}(n) = \bm{W}_{out} \bm{x}(n)$.
The matrices $\bm{W}_{in}$ and $\bm{W}$ are randomly generated and fixed~\cite{Lukosevicius2012}. Only $\bm{W}_{out}$ is trained to minimize \eqref{eq:EPhys}. Following~\cite{Pathak2018a}, each row of $\bm{W}_{in}$ has only one non-zero element, which is randomly drawn from a uniform distribution over $[-\sigma_{in},\sigma_{in}]$; $\bm{W}$ has an average connectivity $\langle d \rangle$, whose non-zero elements are drawn from a uniform distribution over the interval $[-1,1]$; and $\bm{W}$ is scaled such that its largest eigenvalue is $\Lambda\leq 1$, which ensures the Echo State Property \cite{Lukosevicius2012}.  
The training of the PI-ESN is achieved in two steps. First, the network is initialized by an output matrix, $\bm{W}_{out}$, that minimizes a data-only cost functional $E_{tot}^{NP} = E_d + \gamma || \bm{w}_{out,i} ||^2
\label{eq:err_data}$, where $\gamma$ is a Thikonov regularization factor and $\bm{w}_{out,i}$ denotes the $i$-th row of $\bm{W}_{out}$. This is the output matrix of  the conventional ESN~\cite{Pathak2018}. 
Second, the physical error~\eqref{eq:EPhys} is minimized with the L-BFGS method~\cite{Byrd1995}, which is a quasi-Newton optimization algorithm. 

\begin{figure}[!ht]
    \centering
    \includegraphics[width=8.6cm]{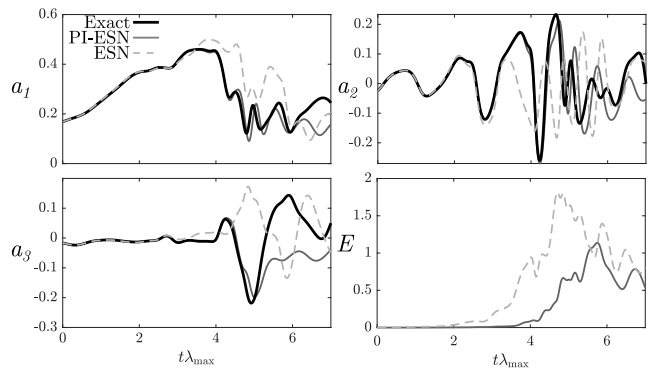}
    \caption{Evolution of $a_1$, $a_2$, $a_3$ during the extreme event of Fig. \ref{fig:MFE_evol}: exact evolution (solid black line), PI-ESN prediction (solid grey line), and conventional ESN prediction (dashed grey line) with reservoirs of $N_x=3000$ neurons. The error of the PI-ESN and ESN predictions is $E$.   }
    \label{fig:MFE_ESN_comp}
\end{figure}

\begin{figure}[!ht]
    \centering
    \includegraphics[width=8.6cm]{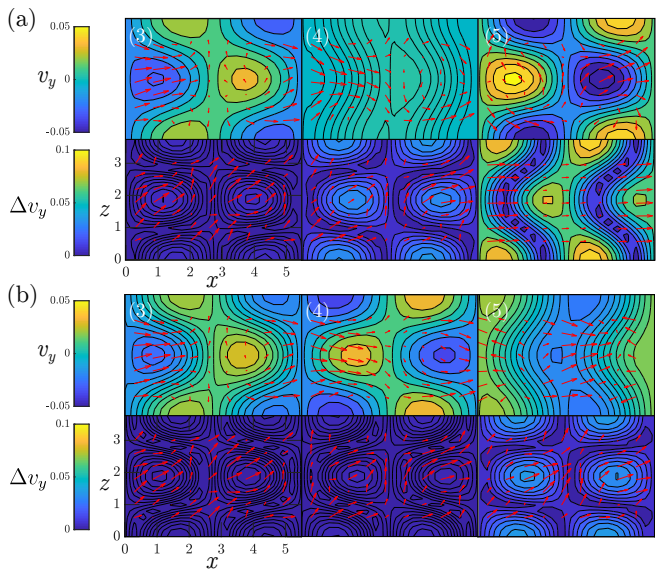}
    \caption{Evolution of the velocity field (top rows) and the normalized error (bottom rows) in the velocity field in the mid-$y$ plane at the same time instants as panels (3)-(5) of Fig. \ref{fig:MFE_ESN_comp}.  Predictions from (a) the conventional ESN and (b) the PI-ESN. The arrows indicate the in-plane velocity ($x$-$z$ directions) and the coloured contour indicates the out-of-plane velocity. The panels correspond to $t^+ = t\lambda_{\max} \approx 2.14,~3.21,~4.27$ in Fig. \ref{fig:MFE_ESN_comp}. 
    }
    \label{fig:MFE_ESN_comp_2D}
\end{figure}

{\it Results.--}
A grid search (see Supplementary Material) provides the following set of hyperparameters for tuning, which perform satisfactorily in the range of $N_x=[500, 3000]$ neurons: $\Lambda = 0.9$, $\sigma_{in}=1.0$, $\langle d \rangle = 3$, $\gamma = 10^{-6}$. Only $t=2500$ time units (equivalent to $t^+\approx 61$) in the window $t=[11500, 14000]$ (equivalent to $t^+\approx[280, 341]$ in the grey box of Fig.~\ref{fig:MFE_evol}) are used for training. The data beyond this time window is used for validation only. We use $N_p=5000$ collocation points (equivalent to $t=1250$ or $t^+\approx30.5$), which provide  a sufficient number of predictions beyond the training data  with a relatively low computational time. %
Fig. \ref{fig:MFE_ESN_comp} shows the evolution of three representative modes' amplitudes during the extreme event in the dashed red box in the top panel of  Fig.~\ref{fig:MFE_evol}. The PI-ESN solution (solid grey line) and the conventional ESN solution (dashed grey line) are computed with a reservoir of $N_x=3000$ units, and they are compared against the exact solution from numerical integration (solid black line). The normalized error  between the exact evolution and the machine predictions is computed as $E(n) = \left(||\bm{y}(n) - \widehat{\bm{y}}(n) ||\right)/\sqrt{\frac{1}{N_t}\sum_{n=1}^{N_t} || \bm{y}(n)||}$~\footnote{The denominator of the error cannot be zero because the fixed point of the MFE model has $||\bm{y}|| = 1$ and  the system has unsteady chaotic oscillations.}.
Although the same training data is used for both the PI-ESN and the conventional ESN, the PI-ESN has a significantly higher extrapolation-in-time capability than the conventional ESN. To compare the performances, we define the predictability horizon as the time required for $E\geq0.2$ from the same initial condition. The predictability horizon of the PI-ESN is $\approx 2$ Lyapunov times longer than the predictability horizon of the conventional ESN. This significant improvement is achieved by enforcing the prior physical knowledge of the turbulent flow, whose evolution has to uncompromisingly fulfil the momentum and mass conservation laws.  
%
As shown in Fig.~\ref{fig:MFE_ESN_comp}, until $t^+\approx 2.14$, both ESN and PI-ESN accurately predict the flow evolution. The predicted solution from the conventional ESN starts  diverging from the exact evolution at $t^+ \approx 3.21$, which leads to a completely different solution during the extreme event. On the other hand, the PI-ESN is able to time-accurately predict the occurrence and the amplitude of the extreme event. After that the event has occurred, the solution diverges because the butterfly effect is significant. 
The turbulent velocity fields predicted by the conventional ESN and PI-ESN are shown in Fig.~\ref{fig:MFE_ESN_comp_2D}a,b, respectively, which are evaluated at the same times as the exact solution in panels (3)-(5) of Fig. \ref{fig:MFE_evol}.  The bottom rows of Fig. \ref{fig:MFE_ESN_comp_2D}a,b show the normalized absolute error between the predicted velocity field and the exact velocity field. 
The discrepancy in the turbulent velocity field is mainly due to the error on the prediction of the downstream vortex mode, $a_3$ (Fig. \ref{fig:MFE_ESN_comp}). On one hand, because no physical knowledge is constrained in the conventional ESN, the sign and amplitude of  $a_3$ are incorrectly predicted, which means that the out-of-plane velocity evolves in the opposite direction of the exact solution. On the other hand, the PI-ESN is able to predict satisfactorily  both the in-plane velocity and the out-of-plane velocity during the extreme event.

\begin{figure}[!ht]
    \centering
    \includegraphics[width=6cm]{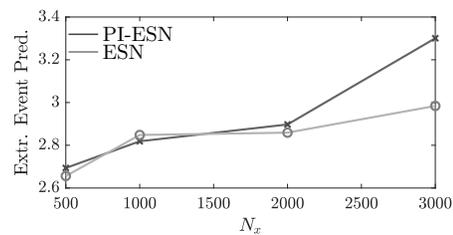}
    \caption{Comparison of the average predictability horizons of the PI-ESN and ESN for all the extreme events in the dataset.}
    \label{fig:MFE_comp_predRE}
\end{figure}

{\it Robustness.--}
To quantitatively assess the robustness of the results, we compute the average predictability horizon of the machines with no further training.  We follow the following steps: 
(i) by inspection of Fig.~\ref{fig:MFE_evol}, we define events as extreme when their kinetic energy is $k\geq0.1$; 
(ii) we identify the times when all the extreme events start in the dataset of Fig.~\ref{fig:MFE_evol}; 
(iii) for each time, the exact initial condition at $t^+ \approx 0.61$ just before the time instant in which the extreme events starts is inputted in the PI-ESN and ESN; 
(iv) the machines are evolved to provide the prediction; and 
(v) the predictability time is computed by averaging over all the extreme events in the dataset. The results are parameterized with the size of the reservoir, $N_x$ (Fig. \ref{fig:MFE_comp_predRE}).
On one hand, for small reservoirs ($N_x\lesssim2000$), the performances of the ESN and PI-ESN are comparable. This means that the performance is more sensitive to the data cost functional, $E_d$, than the physical error, $E_p$.
On the other hand, for larger reservoirs ($N_x\gtrsim2000$), the physical knowledge is fully exploited by the PI-ESN. This means that the performance becomes markedly sensitive to the physical error, $E_p$. This results in a improvement in the average predictability of $\approx 1.5$ Lyapunov times. Because an extreme event takes $\approx1$ Lyapunov time on average, the improved predictability time of the PI-ESN is key to the time-accurate prediction of the occurrence and amplitude of the extreme events. 
%
%
%

{\it Conclusions and perspectives.--} 
The combination of empirical modelling -- based on reservoir computing -- with physical modelling -- based on conservation laws -- enables the time-accurate prediction of extreme events in a turbulent flow. We have compared the performance of a physics-informed echo state network (PI-ESN) and a conventional echo state network (ESN). The difference between the two networks is that the former is a physics-aware machine, whereas the latter is a physics-blind machine because it is trained with data only. 
In the PI-ESN, the physical error from the conservation laws is minimized beyond the training data. This brings in crucial information, which can be exploited in two ways:
(i) with the same amount of available data, the PI-ESN solution is accurate for a longer time than the conventional ESN solution; or %
(ii) less data is required to obtain the same accuracy as the conventional ESN. 
 In this Letter, we have taken advantage of property (i) for the prediction of extreme events in a turbulent flow. 
In future applications of physics-aware machines, the approach should be extended to higher dimensional dynamical systems, such as three-dimensional turbulent flows computed by high-fidelity simulations. This is challenging because the reservoir increases the degrees of freedom of the simulation, which can be aided by both massive GPU computations and nonlinear model reduction.  
Second, the approach should be extended to tackle dynamical systems that contain stochastic processes. This will be useful to filter out the noise from experimental data to use in the training. These aspects are currently investigated in other studies.
In conclusion, this study opens up new possibilities for enhancing synergistically data-driven methods with physical knowledge for the accurate prediction of extreme events in chaotic dynamical systems. 

\begin{acknowledgments}
The authors acknowledge the support of the TUM Institute for Advanced Study funded by the German Excellence Initiative and the EU 7$^{th}$ Framework Programme n. 291763. L.M.  acknowledges support from the Royal Academy of Engineering Research Fellowships Scheme.
\end{acknowledgments}

\bibliographystyle{apsrev4-2}
\bibliography{library}

\end{document}